# Influence of the Statistical Shift of Fermi Level on the Conductivity Behavior in Microcrystalline Silicon

Sanjay K. Ram*,[a], Satyendra Kumar*,[b] and P. Roca i Cabarrocas[!]

*Department of Physics, Indian Institute of Technology Kanpur, Kanpur-208016, India
[!]LPICM, UMR 7647 - CNRS - Ecole Polytechnique, 91128 Palaiseau Cedex, France

The electrical conductivity behavior of highly crystallized undoped hydrogenated microcrystalline silicon (µc-Si:H) films having different microstructures was studied. The dark conductivity is seen to follow Meyer Neldel rule (MNR) in some films and anti MNR in others, depending on the details of microstructural attributes and corresponding changes in the effective density of states distributions. A band tail transport and statistical shift of Fermi level are used to explain the origin of MNR as well as anti-MNR in our samples. We present the evidence of anti MNR in the various experimental transport data of µc-Si:H materials reported in literature and analyze these data together with ours to show the consistency and physical plausibility of statistical shift model. The calculated MNR parameters and other significant material parameters derived therefrom are tenable for a wide microstructural range of the µc-Si:H system.

PACS numbers: 73.50.–h, 73.61.–r, 73.61.Jc, 73.63.–b, 68.55.Jk, 61.72.Mm, 71.23.–k

## I. INTRODUCTION

Measurement of electrical conductivity as a function of temperature is a tool conventionally used to study the electrical transport behavior of disordered systems where an exponential relationship is observed between the conductivity prefactor ($\sigma_0$) and the conductivity activation energy ($E_a$), known as the Meyer-Neldel Rule (MNR or the compensation law).[1] The relationship is expressed as:

$$\sigma_0 = \sigma_{00} e^{GE_a} \qquad (1)$$

where $G$ and $\sigma_{00}$ are called MN parameters. Often $G^{-1}$ is denoted as $E_{MN}$, the Meyer-Neldel characteristic energy. MNR is a phenomenon seen in many thermally activated processes, including electronic conduction in amorphous silicon (a-Si:H). However, the microscopic origin of the MNR and the physical meaning of $G$ are still a topic of discussion. Various theories have been put forward to explain the observed MNR in a-Si:H,[2,3] the most popular among these being the model invoking a statistical shift of Fermi-level ($E_f$) with temperature.[4,5,6,7,8,9,10] In contrast to the homogeneous a-Si:H, hydrogenated microcrystalline silicon (µc-Si:H) is a heterogeneous material consisting of a microcrystallites phase that is comprised of grains which conglomerate to form columns, and amorphous (or disordered) phase and voids populating the inter-grain and inter-columnar boundaries.[11,12,13] MNR has been reported in doped µc-Si:H films with MNR parameters similar to those obtained in a-Si:H,[14] which was explained in terms of statistical shift model analogous to a-Si:H and this formed the basis of several reports that treated the transport in µc-Si:H at par with a-Si:H. This led to the general belief that while the optical properties of µc-Si:H are governed by the crystalline component, the electrical transport is still controlled by the amorphous silicon phase.

Any analogy that does exist between µc-Si:H and a-Si:H materials gets somewhat undermined in the newer high mobility µc-Si:H materials having complete crystallization from the beginning of film growth.[15,16,17,18,19] In such a material the absence of an amorphous phase gives rise to mechanisms and routes of electrical transport different from our conventional understanding of relationship between electrical transport and variation in crystallinity.[19] Apart from MNR, another interesting and important phenomenon, the anti MNR, has been reported in heavily doped µc-Si:H[20,21,22] and heterogeneous Si (het-Si) thin film transistors (TFTs).[23] A negative value of MN energy $E_{MN}$, is seen in case of anti-MNR. This phenomenon has been explained by the $E_f$ moving deep into the band tail. Anti-MNR is not seen in a-Si:H, the accepted reason being that it is difficult to dope a-Si:H heavily enough to move $E_f$ deeply into the tail DOS, due to disorder induced broadening of the tail state distribution.

Therefore, the observation of MNR and anti MNR in electrical transport behavior of µc-Si:H draws our attention towards the basic physics, in terms of both the origin and significance of these relationships. In spite of the immense potential of high efficiencies and large area deposition capabilities shown by µc-Si:H in semiconductor technology, especially in photovoltaics[24,25] and TFTs,[26] the understanding of its transport properties is impeded by these lacunae. Different conduction mechanisms and paths have been invoked to explain the electrical transport behavior in µc-Si:H, deriving information that correlate only some microstructural features and mechanisms. Azulay *et al.* have discussed such models.[19] However, the present approach does not provide an insight into what occurs at the electronic level, the knowledge of which could present a more unified picture of the electronic transport. The inherent microstructural complexities, the consequent intricacy of electrical transport behavior of this heterogeneous material and the lack of knowledge regarding a reproducible relationship between the two continues to remain an obstacle to a clear view of the composite picture. The unavailability of an effective density of states (DOS) map of the heterogeneous µc-Si:H system contributes to the problem.

---

[a] Corresponding author. E-mail address: skram@iitk.ac.in, sanjayk.ram@gmail.com
[b] satyen@iitk.ac.in

In this paper, we present the results of dark conductivity ($\sigma_d$) measurements (above room temperature) conducted on a large number of well-characterized µc-Si:H samples having a broad range of microstructural features. In particular, fully crystallized µc-Si:H films with no amorphous tissue have been studied. In Sec. II, the experimental details are described. In Sec. III, we present the results of our measurements, which evince the occurrence of both MNR and anti-MNR in undoped µc-Si:H. We have discussed these findings in the context of statistical shift models in Sec. IV, where we show that these phenomena are intricately linked to the underlying microstructure and the corresponding DOS features of the material.[27] Further, a number of published temperature dependent conductivity data has been analyzed in a consistent framework. The conclusions are presented in Sec. V.

## II. EXPERIMENT

The undoped µc-Si:H films were deposited at low substrate temperature ($T_s \leq 200°C$) in a parallel-plate glow discharge plasma enhanced chemical vapor deposition system operating at a standard rf frequency of 13.56 MHz, using high purity $SiF_4$, Ar and $H_2$ as feed gases. Different microstructural series of samples were created by systematically varying gas flow ratios ($R = SiF_4/H_2$) or $T_s$ (100-250°C) for samples having different thicknesses (≈50-1200nm). We employed bifacial Raman scattering (RS), spectroscopic ellipsometry (SE), X-ray diffraction (XRD), and atomic force microscopy (AFM) for structural investigations. Total crystalline volume fraction in the bulk of all films was more than 90% from the beginning of the growth, with the rest being voids, as revealed by RS and SE analyses. A thin (≈20 nm) film-substrate interface incubation layer is detected by both SE and RS (substrate side), which is less crystallized, having some amorphous content. The fractional composition of the films educed from SE data showed crystallite grains of two distinct sizes,[18] which was corroborated by the deconvolution of RS profiles using a bimodal size distribution[28,29] of large crystallite grains (LG ≈ 70-80nm) and small crystallite grains (SG ≈ 6-7nm). The XRD results have demonstrated the LG and SG to be having different orientations. However, there is a significant variation in the percentage fraction of the constituent LG ($F_{cl}$) and SG ($F_{cf}$) with film growth. The details of the microstructural findings and modeling method used for incorporation of the bimodal crystallite size distribution are being reported elsewhere. The evolution of conglomeration of crystallites under different deposition conditions was demonstrated by AFM results.[18] Preferential orientation in (400) and (220) directions was achieved by optimizing the deposition conditions leading to smooth top surfaces (surface roughness < 3nm), indicating device quality material. Many of the µc-Si:H films used in this study have been characterized by the time resolved microwave conductivity (TRMC) measurements as well. Coplanar $\sigma_d(T)$ measurements were carried out from 300K to 450K on these well-characterized annealed samples having a variety of film thicknesses and microstructures, and studied in context of deposition parameters.[27] At above room temperature, $\sigma_d(T)$ of all the µc-Si:H films having different microstructures, prepared under different deposition conditions, follows Arrhenius type thermally activated behavior:

$$\sigma_d = \sigma_0 e^{-E_a/kT} \qquad (2)$$

## III. RESULTS

A systematic study and analysis of the results of the large number and variety of samples in this study necessitates organizing them into some classes having common attributes. Therefore, on the basis of structural investigations of the µc-Si:H films at various stages of growth and under different growth conditions, we have segregated out the unique features of microstructure and growth type present in the varieties of films, with respect to the correlative coplanar electrical transport properties and classified them into three types: *A*, *B* and *C*. Since this classification is fundamental to this work, we first describe it briefly. This classification has been dealt with in detail in a work being reported elsewhere.

The first issue that arises when we embark on such a classification is the choice of microstructural parameter that can be correlated to the observed electrical transport behavior. Deposition parameters cannot be a rational choice as they only have an indirect causal link to the electrical properties through their primary effect on the microstructure of material. In the beginning, we had attempted to correlate the classification to thickness, as a particular type of electrical transport behavior appeared to exist over a certain range of thickness (*type-A* up to

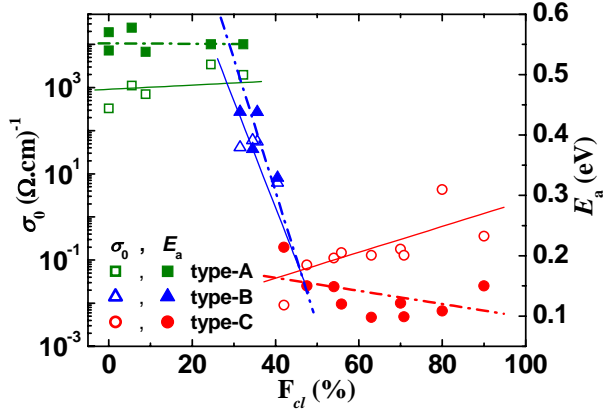

FIG. 1. Variation of $\sigma_0$ and $E_a$ of $\mu$c-Si:H samples (*types*: *A*, *B* and *C*) with $F_{cl}$.

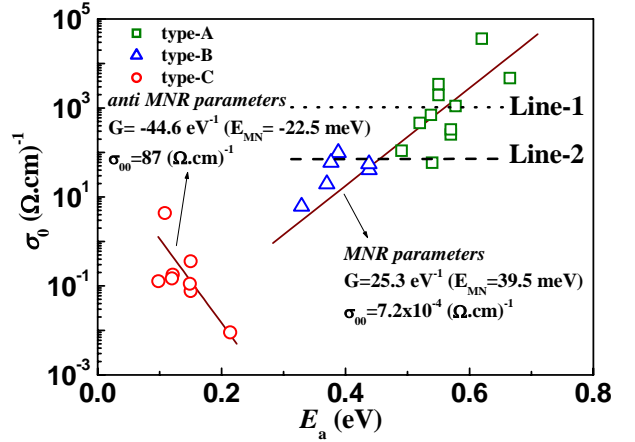

FIG. 2. Correlation between $\sigma_0$ and $E_a$ in undoped $\mu$c-Si:H samples (*types*: *A*, *B* and *C*). The samples of *types-A* and *B* follow MNR while *type-C* material shows anti MNR behavior with parameters as shown in the graph. In the MNR region the dotted line (Line-1) indicates the possible position of $\sigma_0$ where $\gamma_f \approx 0$ and the dashed line (Line-2) where $\gamma_f \approx \gamma_c$.

≈350–400nm, *type-B* in the range 400–900 nm, and *type-C* above 900nm). However, the observed electrical transport behaviors could not be explained solely based on thickness, and each time we had to resort to a correlation between thickness and $F_{cl}$, which led us to reconsider our choice of parameter. An extensive analysis showed that rather than film thickness or any other deposition parameter, it is the $F_{cl}$ which reflects the microstructural and morphological stage of the film, and correlates well with the electrical transport behavior.[27,30] We have few samples in this study having thicknesses that could belong to a certain type in a classification system based on thickness, but the observed electrical transport behaviors were not correlative with the 'thickness zone' category in these samples. The electrical transport behaviors were rather correlative with the $F_{cl}$ range each sample falls in and the classification based on such $F_{cl}$ values is physically more rational.

The classification is depicted in Fig. 1 where the variation of $\sigma_0$ and $E_a$ of the three types of films with $F_{cl}$ is shown. To summarize this classification, the *type-A* films have small grains, low amount of conglomeration (without column formation), and high density of inter-grain boundary regions containing disordered phase. In this type, $F_{cl}$<30%, $\sigma_0$ and $E_a$ are constant [≈10$^3$ ($\Omega$cm)$^{-1}$ and ≈0.55 eV respectively]. The *type-B* films contain a fixed ratio of mixed grains in the bulk. There is a marked morphological variation in these films due to the commencement of conglomeration of grains resulting in column formation, and a moderate amount of disordered phase is present in the columnar boundaries. Here $F_{cl}$ varies from 30% to 45% and there is a sharp drop in $\sigma_0$ [from ≈10$^3$ to 0.1 ($\Omega$cm)$^{-1}$] and $E_a$ (from ≈0.55 to 0.2 eV). The *type-C* $\mu$c-Si:H material is fully crystallized, crystallite conglomerates are densely packed with significant fraction of large crystallites (>50%) and preferential orientation is seen. Here $\sigma_0$ shows a rising trend [from 0.05 to 1 ($\Omega$cm)$^{-1}$] and the fall in $E_a$ is slowed down (from 0.2 to 0.10 eV). The microstructural features that result in such changes in the electrical transport behavior are discussed later.

While a relation between the $F_{cl}$ and electrical transport parameters is evident in the Fig. 1, it is also observed that there is some relation between $\sigma_0$ and $E_a$. Therefore, it would be useful to study the variation of $\sigma_0$ with $E_a$ for each type of material. Fig. 2 shows a semi-logarithmic plot between $\sigma_0$ and the $E_a$ obtained on our samples. The data for *types-A* and *B* are found to fall along the MNR line. We found the values of the MNR parameters, $G \approx 25.3$ eV$^{-1}$ (or $E_{MN} \approx 39.5$ meV) and $\sigma_{00} \approx 7.2 \times 10^{-4}$ ($\Omega$cm)$^{-1}$ from the fit shown in the figure. In contrast, the data for samples of *type-C* shows an inverse linear relationship between logarithmic value of $\sigma_0$ and $E_a$. The correlation between $\sigma_0$ and $E_a$ appears to change sign in this case, known as / demonstrating *anti MNR* and the values of the MNR parameters are: $G \approx -4.6$ eV$^{-1}$ or $E_{MN} \approx -22.5$ meV and $\sigma_{00} \approx 86.8$ ($\Omega$cm)$^{-1}$.

## IV. DISCUSSION

In order to understand the MNR parameters, anti MNR phenomenon and their significance, we discuss here the applicability of the existing statistical shift model developed for *a*-Si:H in explaining both MNR and anti MNR behavior in the conductivity of $\mu$c-Si:H. In a disordered semiconductor, the DOS distribution may not be symmetrical with respect to the band center due to tailing of localized states at the band edges as well as defect states in the gap. Therefore, the position of $E_f$ is determined by the shape of the DOS, i.e. at $T=0K$, $E_f$ may be at some position other than the mid-gap. However, as the temperature increases, electrons and holes are thermally excited to states at the band edges, and charge neutrality condition requires a statistical shift of $E_f$ towards mid-gap. Therefore, the experimentally obtained $\sigma_0$ contains terms arising from the two effects. The first comes from the statistical shift of $E_f$ and the second involves a temperature dependent shift of the band edges, i.e., of conduction and valence band edges, $E_c$ and $E_v$.[4] According to Mott, one can express the conductivity expression as:[31]

$$\sigma_d(T) = \sigma_M \exp(-(E_c - E_f)/kT)) \qquad (3)$$

where $\sigma_M$ is minimum metallic conductivity. $E_c$ and $E_f$ are both dependent on temperature. Approximating the temperature shift of $E_c$ and $E_f$ to be linear functions with the slopes $\gamma_c$ and $\gamma_f$ respectively, we get

$$E_c(T) = E_c^0 - \gamma_c T \quad \text{and} \quad E_f(T) = E_f^0 - \gamma_f T \qquad (4)$$

where $E_c^0$, $E_f^0$ are the positions of $E_c$ and $E_f$ at $T=0K$. After inserting Eq. (4) into Eq.(3), we get Eq.(2) with

$$E_a = E_c^0 - E_f^0 \qquad (5)$$

$$\sigma_0 = \sigma_M \exp[(\gamma_c - \gamma_f)/k] \qquad (6)$$

The band shifts are taken relative to midgap. They are positive when $E_c$ and $E_f$ move towards midgap.

Normal MNR in $\mu$c-Si:H has been generally understood using the above calculations, but the anti MNR behavior is rather less elucidated. Most workers[22,23] have attributed the anti MNR behavior observed in doped $\mu$c-Si:H material to the model implicating energy band (EB) diagram of crystalline silicon (c-Si) and a-Si:H interface as proposed by Lucovsky and Overhof (referred to as LO in this work).[20] According to this model, anti MNR can be observed only in a degenerate case when very heavy doping of the $\mu$c-Si:H material causes $E_f$ to move above $E_c$ in the crystalline phase and consequently $E_f$ can move deeply into the tail states in the disordered region. In explaining anti MNR behavior on the basis of this EB diagram, equal band edge discontinuities at both ends of c-Si and a-Si:H interface were assumed.[20] A similar argument was also given by Meiling et al.[23] in their study of intrinsic het-Si TFT. But it is not a clear-cut task to calculate the EB diagram of the interface because the band edge discontinuities are not really well established, as some studies have claimed the band edge discontinuities between crystalline grains and amorphous tissue regions to be most pronounced in the conduction band (CB),[32] while others have attributed the discontinuity mainly to the valence band (VB).[33,34] However, recent findings support the latter contention.[35]

The loss of linearity of the relationship between $E_a$ and log$\sigma_0$ (i.e., anti MNR) in a-Si:H has been observed for both very high and very low values of $E_a$ in theoretical studies where calculations based on statistical shift model were applied.[5,6,7,8,9] In these studies, the occurrence of anti MNR at high $E_a$ in a-Si:H has been dealt with employing both experimental and theoretical approaches. However, the lack of experimental evidence for anti MNR at low $E_a$ in a-Si:H was probably the reason why it was not elaborated theoretically in much detail.[36] An important insight these studies offer is that the occurrence of anti MNR depends on the DOS distributions and the position of the $E_f$ in the a-Si:H material. These studies are relevant for $\mu$c-Si:H, because the same conditions that give rise to anti MNR in a-Si:H at low $E_a$, may be applicable in the $\mu$c-Si:H system as well. The linear relation between log$\sigma_0$ and $E_a$ in a-Si:H is obtained only if $E_f$ lies in the CB tail or close to the minimum of DOS. However, there may exist a combination of factors, which can serve to either augment or diminish the anti MNR effect. Like, when $E_f$ approaches the boundaries, if there is a flat DOS spectrum near the edge, the statistical shift of $E_f$ will diminish along with a decrease in the temperature derivative of $E_f$. This causes anti MNR behavior for small and large $E_a$ regions.[8,9] However, such a deviation (anti MNR effect) may diminish if there is a jump in the DOS profile present at the edge of the steep tail.[9] Another situation is where the anti MNR is more pronounced at values of $E_a$ on the lower side, when the DOS value at the minimum reduces.[9] Such a reduction in DOS value at the minimum, where exponential CB and negatively charged dangling bond (DB$^-$) tails meet, has been observed in the case of n-type doping of a-Si:H.[37,38,39] Here DOS of DB$^-$ band increases with increased doping level. The failure of MNR at higher $E_a$ side can be seen when $E_f$ lies far in the tail of the DB$^-$ states or in intrinsic materials where DOS at the mid gap is almost flat, due to which the temperature derivative of $E_f$ will be very little or almost zero.[9]

An important aspect of understanding electronic transport is the actual conduction route. All the above concepts hold true for $\mu$c-Si:H only if a band tail transport exists. There has been a lack of a consensus on what the current flow pathway is in fully crystallized undoped $\mu$c-Si:H. One school of thought propounds the idea that conduction takes place through individual crystallites or aggregates thereof (columns).[40,41] In discussing transport mechanisms in $\mu$c-Si:H we need to distinguish between highly doped samples and undoped or unintentionally doped samples. In heavily doped $\mu$c-Si:H material current route follows through crystallites/ columns and transport properties can be understood by well established grain boundary trapping (GBT) models. However, an alternate viewpoint that is also borne out by recent experimental evidence suggests a dominant role of the disordered silicon tissue of the boundaries encapsulating the crystallite columns in electrical transport in fully crystalline single phase undoped $\mu$c-Si:H material.[19]

Therefore, in order to understand the transport data of our $\mu$c-Si:H samples, we need to consider the data in context of not only the respective sample microstructures, but also in respect to the above-discussed theoretical background and the possible current routes in the material. Though many studies have reported on electrical transport properties of highly crystalline $\mu$c-Si:H material, it is noteworthy that in these cases, such high crystallinity (>80%) is achieved around the time when column formation is also complete during the film growth process.[42,43] In poorly crystalline $\mu$c-Si:H material, where an interconnected network of boundary tissue has not formed, transport is considered to take place through the amorphous matrix.[19] Our material is somewhat different, as it achieves full crystallization from the beginning of the growth due to the use of SiF$_4$ in deposition, which is especially relevant in the *type-A* material, where there is almost complete crystallization, without an amorphous matrix, though the column formation has not started.[18,27] The *type-A* material consists mainly of SG with an increased number of SG boundaries. Therefore, the question of formation of potential barrier (i.e., transport through crystallites) does not arise because the large number of defect/trap sites compared to free electrons and small size of crystallites will result in a depletion width that is sufficiently large to become greater than the grain size, causing the entire grain to be depleted.[40] Therefore, the transport will be governed by the band tail transport. Corroboratively, a look at Fig. 1 shows that in *type-A* material, $E_a$ becomes nearly saturated (≈0.55eV) and $\sigma_o$ reaches ≈10$^3$ ($\Omega$cm)$^{-1}$. This means the $E_f$ is lying in the gap where the DOS does not vary much and there is a minimal movement of $E_f$, or $\gamma_f \approx 0$.[6,7,8] We have indicated this possible position of $\sigma_0$ where $\gamma_f = 0$ in Fig. 2 by a dotted line (Line-1). The initial data points shown in Fig. 2 for *type-A* have higher $\sigma_o$ [≈ 10$^4$ ($\Omega$cm)$^{-1}$] and $E_a$ (≈0.66eV), because of a shift in $E_c$ and/or a negative value of $\gamma_f$, as happens in *a*-Si:H for $E_a$ towards the higher side.[6,7,8]

In *type-B* material, many morphological changes are occurring during the film growth, column formation has commenced, and there is a change in the transport routes. It is a crossover region from *type-A* to *type-C* that shows large variations in $\sigma_0$ and its $E_a$ as seen in Fig. 1. The improvement in film microstructure leads to a delocalization of the tail states causing the $E_f$ to move towards the band edges, closer to the current path at $E_c$. The statistical shift $\gamma_f$, depends on the temperature and the initial position of $E_f$, and when the $E_f$ is closer to any of the tail states and the tail states are steep, $\gamma_f$ is rapid and marked. In fact, any $\mu$c-Si:H material with such microstructural attributes as in *type-B* material, will show a rapid change in $\gamma_f$. In Fig. 2, the transition between *type-A* and *type-B* materials shows a few data points somewhat scattered around the MNR line, belonging to both the types, which show a more or less constant $\sigma_o$ [70-90 ($\Omega$cm)$^{-1}$] with the fall in $E_a$ (0.54-0.40 eV), indicating that the temperature shift of $E_f$ and that of the CB have become equal, canceling each other out (i.e., $\gamma_f \approx \gamma_c$).[6,7,8] We have depicted the possible position of $\sigma_0$ where such a situation can occur in Fig. 2 by a dashed line (Line-2). In this case, the $E_f$ is pinned near the minimum of the DOS between the exponential CBT and the tail of the defect states (DB$^-$).[6,7,8] With increasing crystallinity and/or improvement in the microstructure, the minimum shifts towards $E_c$ leading to a decrease of $E_a$.

In *type-C* material, one can erroneously assume the apparent low values of $E_a$ to be GB barrier height formed at the interface between neighboring crystallites/columns, and the appearance of reduction in $E_a$ to be a reduction in barrier height with film growth/ increasing $F_{cl}$ (as shown in Fig. 1), in a manner similar to the situation when the mobility-barrier height variation is seen to match the conductivity-$E_a$ with increase in doping.[44] The calculated values of free electron concentrations (from $\mu_{TRMC}$, $\sigma_d$ and $E_a$ data) do not suggest the possibility of unintentional doping achieving such a high value of background doping concentration. So the unintentional doping resulting in degeneracy is not possible either. Therefore, the EB model as suggested by Lucovsky *et al.* seems inapplicable to our undoped $\mu$c-Si:H case, though the value of $E_{MN}$ here is close to the value reported in heavily doped $\mu$c-Si:H (-20meV).[20] Also, in a degenerate case, the conductivity behavior of polycrystalline material is found to exhibit a $T^2$ dependence of $\sigma_d$,[45] which is not so in our material. In *type-C* $\mu$c-Si:H material, a higher F$_{cl}$ and large size of columns (>300nm) result in less columnar boundaries, and therefore less defects associated with the boundaries. In addition, a well-established conducting network of such interconnected boundaries is formed in this material, which results in higher conductivity (rise in $\sigma_0$). Considering transport through the encapsulating disordered tissue, a band tail transport is mandatory. The large columnar microstructure results in a long range ordering which is sufficient to delocalize an appreciable range of states in the tail state distribution. In addition, higher density of available free carriers and low value of defect density can cause a large increase in DB$^-$ density together with a decrease in positively charged dangling bond (DB$^+$) states in the gap, which results in a lower DOS near the CB edge and can create a possibility of a steeper CB tail. In this situation, if $E_f$ is lying in the plateau region of the DOS, it may create an anti MNR situation.

Our experimental and modeling study of phototransport properties of these three types of $\mu$c-Si:H materials had evinced a fundamental change taking place in the DOS distributions along with change in

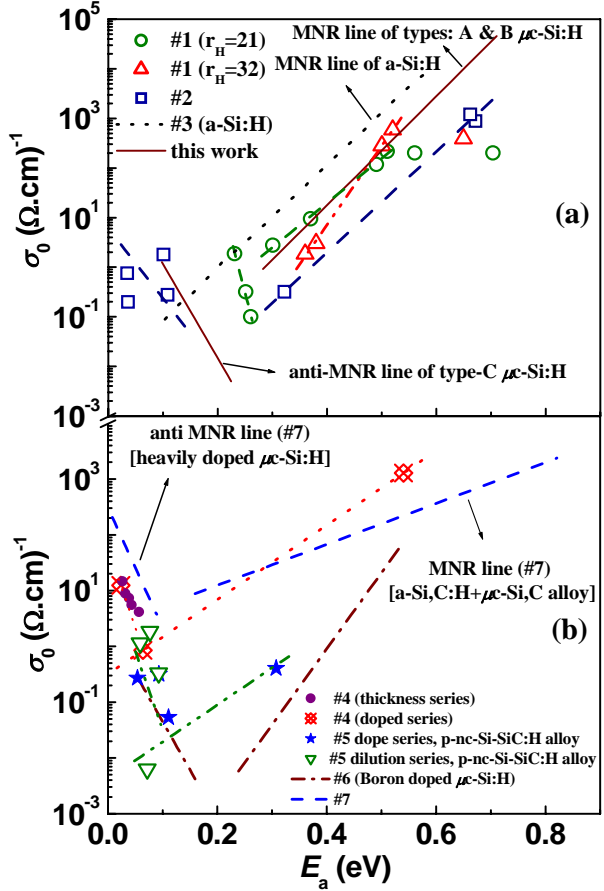

FIG. 3. Plot of $\sigma_o$ as a function of $E_a$ for data of heavily doped, doped and undoped $\mu$c-Si:H, $p$-nc-Si-SiC:H alloy, a-Si,C:H + $\mu$c-Si,C alloy, and $a$-Si:H obtained from literature. (a) case #1- Ref. 42, case #2- Ref. 50, case #3- Ref. 49; (b) Similar data for case #4- Ref. 51, case #5- Ref. 53 and 54, case #6- Ref. 21, case #7- Ref. 20. The solid line represents our data.

microstructure and electrical transport properties.[46,47,48]

Therefore, as we go from one type of material to another, a change occurs in the DOS of the transport path elements as well, and this DOS cannot be assumed to be similar to that of $a$-Si:H or c-Si, because the unique film microstructure will have an effect on it. Our proposed effective DOS distributions are different for the three types of $\mu$c-Si:H materials, and exhibit structured band tails: a sharper, shallow tail originating from grain boundary defects and another less steeper deep tail associated with the defects in the columnar boundary regions, both of which have an exponential distribution.[46,47,48] Therefore, considering all the above reasons, a band tail transport presents an experimentally and theoretically consistent picture of the transport in our $\mu$c-Si:H materials.

Anti MNR behavior in undoped $\mu$c-Si:H has not been explicitly reported in literature, to the knowledge of the authors. When we suggest that the anti MNR behavior observed in our *type-C* material has not anomalously arisen from unintentional doping (which has been ruled out in discussions above), and is explainable solely on the basis of DOS features, it follows naturally that similar reproducible anti MNR relationship should be found in similar type of undoped $\mu$c-Si:H material. With this aim, we have analyzed the transport data of highly crystalline undoped $\mu$c-Si:H material reported in literature to see whether we can find any signature of anti MNR in those data. Many of these works were not concerned with the study of conductivity behavior in the context of MNR or anti MNR; and where the relevant parameter values were not given explicitly, we have determined them by analyzing the data published in these papers for the presence of MNR and anti MNR in their materials. Fig. 3(a) shows the plot between $\sigma_o$ and $E_a$ of undoped $\mu$c-Si:H material (cases #1), and $p$-doped $\mu$c-Si:H material (case #2), the values of which has been derived from data in literature. Also shown in this figure are the reported results on $a$-Si:H (typical MNR, case #3, Ref. 49). For the purpose of comparability, our data (*types- A, B* and *C* $\mu$c-Si:H materials) is shown by the solid line, which is the same fitted line shown in Fig 2. Figure 3(b) (cases #4 and 5) shows the plot between $\sigma_o$ and $E_a$ for doped $\mu$c-Si:H material reported in literature. The reported results of MNR and anti MNR in doped $\mu$c-Si:H from Flückiger et al. (case#6) and those of anti MNR in heavily doped $\mu$c-Si:H from Lucovsky et al.[20] (case#7) are also shown in the Fig. 3(b). Though MNR and anti MNR behaviors are visible on even a cursory look at the figures, we will now take up each data and examine it in relation to the sample microstructure as reported in the respective paper. The various data that fall in the MNR region show a variation in the MNR parameters $G$ and $\sigma_{00}$, the values of which are mentioned in the figure captions. $G$ is a strong function of the DOS in the mobility gap and thus of the position of $E_f$. It varies from one DOS to another and if the slope of DOS around the minima becomes steeper, $G$ increases.

First, we take up the *case#1* [Fig. 3 (a), data of Kočka et al.[42]]. Here the samples deposited at $r_H$ (H$_2$ / SiH$_4$) =21 (4.5% dilution) have the thickness 0.07 to 4.7 $\mu$m and those at $r_H$ =32 (3% dilution) have thickness 0.1 to 1.7 $\mu$m. The electrical transport data of these samples show MNR behavior and the $G$ of these two series of samples of $r_H$ = 21 and 32 are 20 and 36 eV$^{-1}$ respectively, which signifies that the CBT should be steeper for the latter case. The samples of $r_H$ =21 series that are thicker than 1$\mu$m show anti MNR, here the crystallinity has reached a constant value of ≈90%, and the material has a densely packed columnar microstructure.

The second data of Collins et al.[50] [Fig. 3 (a), *case#2*] incorporates data of $p$-doped $\mu$c-Si:H material having a fixed and low amount of doping, but it is the H$_2$ dilution, and not doping level that is altered to yield different microstructures in this study. Therefore, this data has not been included with the doped samples data shown in Fig. 3(b). The samples of this case belong to three types of materials, namely, amorphous films of low conductivity and high $E_a$ (0.67eV) obtained with $R$ (=H$_2$/ SiH$_4$) <80, mixed-phase (a+μc)-Si:H films ($E_a$ falls with rise in $R$, and $\sigma_d$ increases) obtained with 80<$R$<160, and single-phase μc-Si:H films ($E_a$ is very low ≈0.1eV) obtained with $R$>160 in which nucleation of microcrystallites occurs immediately on the ZnO substrate without an amorphous interlayer. The first two types of materials show MNR behavior, while the data of the material described as single-phase μc-Si:H exhibits anti MNR behavior.

In *case#4* [Fig. 3 (b), He *et al.*[51]], the data consists of two series. The first series consists of doped samples (starting with an undoped sample, and progressively increasing the doping level) having a constant thickness. The second series consists of doped samples with systematically reduced thicknesses, but fixed doping. In the first series, as we proceed from the undoped sample to the sample with the highest doping, a change from MNR to anti MNR behavior is seen. However, in the second series, as the thickness is reduced, $\sigma_0$ falls with an increase in $E_a$, still following the anti MNR line, demonstrating that with microstructural changes, the anti MNR effect reduces. The $G$ value is ≈15.6 eV$^{-1}$, which means the slope of CBT is larger than that of our materials. This might be because in heavily doped case, doping may systematically increase the extent of the tail states, since in *n*-type doping the dominant charged defects are $P_4^+$ and $DB^-$ and the rate of increase of $P_4^+$ defect density is faster than that of $DB^-$.[52] It is evident from the cases # 2 and 4, that the anti MNR behavior is not just an effect of doping, but is an outcome of fundamental microstructural attributes.

Now we consider the *case#5* [Fig. 3 (b), Myong *et al.*[53,54]] which shows the data of hydrogenated boron (B)-doped nc-Si-SiC:H (*p*-nc-Si-SiC:H alloy) material. This is similar to the μc-Si,C alloy material mentioned in Ref. [20], in which the possibility of anti MNR induced by doping was ruled out. The reason stated for this was it is improbable that the Si crystallites can be doped to such a degree that the $E_f$ is driven deep into the band-tail state distribution of the a-Si,C:H phase. The material in case#5 contains nc-Si grains embedded in *a*-SiC:H matrix. We have analyzed the data of two types of samples of this material. In the first type of samples (doping series)[53], boron doping was increased while H$_2$ dilution was kept constant at 20. The second series (dilution series)[54] consists of samples deposited under constant boron doping (B$_2$H$_6$/SiH$_4$ = 1000ppm) and varying H$_2$ dilution (H$_2$/SiH$_4$ = 15–30). The film thickness is constant in both the series. In the doping series, $\sigma_d$ is reported to follow MNR with a very high value of $E_{MN}$ (≈295 meV), which has been explained by thermally activated hopping between neighboring crystallites dominating the carrier transport in $T$ >150 K regime.[53] The possibility of extended-state transport was ruled out in the study.[53] The dilution series has not been studied for MNR behavior in this work.[54] When we plot together the data of samples of both the series, the result surprisingly reveals evidence of anti MNR behavior in addition to MNR behavior [see Fig. 3 (b)]. The only sample with data lying in the MNR region is a completely amorphous highly doped (B$_2$H$_6$/SiH$_4$ = 8000ppm) material with 0% crystallinity. Another sample (B$_2$H$_6$/SiH$_4$ = 8000ppm and crystalline volume fraction = 22.8%) lies in the intermediate region, while rest of the samples have values well within the anti MNR region, and it is the transition region between these two extremes which has been reported in the Ref. [53]. When the $E_{MN}$ is recalculated from this MNR line (after neglecting the anti MNR data), the value of $E_{MN}$ is ≈ 65 meV, which is closer to the typical $E_{MN}$ values reported for MNR line in *a*-Si:H or μc-Si:H materials. This suggests anti MNR behavior is possible in this kind of nc-Si-SiC:H alloy material also, which has not been reported previously, and certainly needs more exploration. It seems likely that the transport mechanism in the MNR regime in this material is not thermally activated hopping[49] as suggested in the Ref. [53]. Both the MNR and anti MNR behavior in this nc-Si-SiC:H alloy material can be better understood by the same statistical shift model which we have used to explain MNR and anti MNR behavior observed in our material.

The *case#6* [Fig. 3 (b), Flückiger *et al.*[21]] consists of the data of a highly crystalline μc-Si:H material studied for compensation doping (*p*-type doping with boron) that shows a transition from anti MNR to MNR behavior with increasing doping. The initial point of anti MNR therefore belongs to the undoped material, and the deposition technique used there (very-high frequency-glow discharge) results in a high crystallinity and such microstructural attributes that can result in the anti MNR behavior.

All the above data suggests anti MNR is possible in undoped μc-Si:H material. Anti MNR has been reported in TFTs incorporating intrinsic heterogeneous Si, but it was explained by applying the LO model.[23] The material in this particular report consisted of cone shaped Si crystals embedded in an amorphous matrix (the cones initiate at the SiO$_2$ surface), and the samples showing anti MNR had fully coalescent crystallites growing perpendicular to the substrate surface, quite similar to our *type-C* material where anti MNR is seen. In the view of all the evidence from our study, it seems possible that statistical shift model can explain the anti MNR in *het*-Si as well. The MNR and anti MNR parameters for all the above-discussed cases are given in Table I.

All the above discussions demonstrate that MNR is valid for the whole class of μc-Si:H materials. The value of MNR parameter $G$ for a particular μc-Si:H material is related to the microstructure and DOS characteristics of that material, although different sets of MNR parameters $G$ and $\sigma_{00}$ values can exist for the materials of the same $\mu$c-Si:H system. We have tried to explain the variation and significance of $G$ in the above discussions. Now, it would be desirable to explore the variation in the values of $\sigma_{00}$ as well, and understand what $\sigma_{00}$ really means in the context of the material. All the above data, including those of ours, has provided us with a number of $G$ and $\sigma_{00}$ values, which should be useful to get an insight into a relationship between the two, similar to the relationship derived by Drusedau et al.[6,7]

Table I. List of MNR and anti MNR parameters for all the cases studied.

| Samples | MNR parameters | | | Anti MNR parameters | | |
|---|---|---|---|---|---|---|
| | $\sigma_{00}$ $(\Omega.cm)^{-1}$ | $G$ $(eV^{-1})$ | $E_{MN}$ (meV) | $\sigma_{00}$ $(\Omega.cm)^{-1}$ | $G$ $(eV^{-1})$ | $E_{MN}$ (meV) |
| This work | | | | | | |
| Type-A&B | $7.2 \times 10^{-4}$ | 25.3 | 39.5 | -- | -- | -- |
| Type-C | -- | -- | -- | 87 | -44.6 | -22.5 |
| Published Data | | | | | | |
| Case#1 ($r_H$=21) | $4 \times 10^{-3}$ | 20.7 | 48.4 | $1.26 \times 10^{10}$ | -97.7 | -10.2 |
| Case#1 ($r_H$=32) | $3.2 \times 10^{-6}$ | 36.6 | 27.3 | -- | -- | -- |
| Case#2 | $1.7 \times 10^{-4}$ | 23.4 | 42.7 | 6 | -32.5 | -30.8 |
| Case#3 | $7.7 \times 10^{-3}$ | 24 | 41.6 | -- | -- | -- |
| Case#4 | 0.32 | 15.4 | 65.1 | 59 | -66.1 | -15.1 |
| Case#5 | $4.2 \times 10^{-3}$ | 15.3 | 65.4 | 21 | -64.9 | -15.4 |
| Case#6 | $3.2 \times 10^{-6}$ | 31.3 | 31.9 | 2.4 | -39.9 | -25.1 |
| Case#7 | 2.3 | 8.5 | 118.3 | 309 | -49.5 | -20.2 |
| Case#8 | 0.5 | 11.8 | 84.5 | -- | -- | -- |
| Case#9 | $7.2 \times 10^{-3}$ | 20 | 50 | -- | -- | -- |

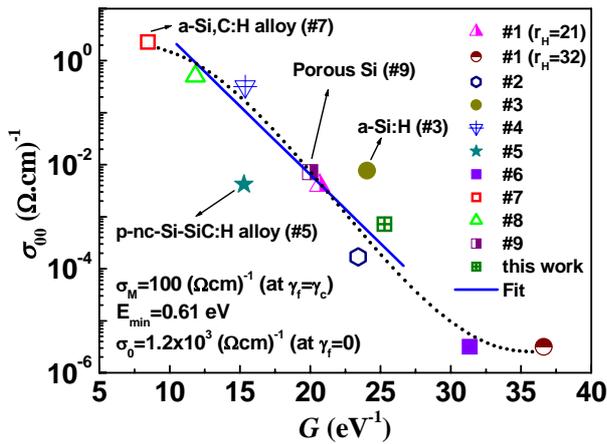

FIG. 4. Relationship between $\sigma_{00}$ and $G$ for cases #1-7 of Fig. 3, case #8- Ref. 55, case #9- Ref. 49 and our data. The solid line shows the fit.

If we compare the Eq. 1 and Eq. 6, we find a relationship between the MNR parameter $\sigma_{00}$ and the fundamental pre-exponential factor or minimum metallic conductivity $\sigma_M$ as:

$$\sigma_{00} = \sigma_M \exp[(\gamma_c - \gamma_f)/k - GE_a] \quad (7)$$

Applying Eq.5 we get

$$\sigma_{00} = \sigma_M \exp[(\gamma_c - \gamma_f)/k - G(E_c^0 - E_f^0)] \quad (8)$$

For a set of samples such as in our case where $\mu$c-Si:H materials possess different DOS, the average statistical shifts of $E_f$ can be assumed to be identical with the temperature coefficient $\gamma_f$ of $E_f$ and can be represented as a function of the position of $E_f$ with an energy $(E_c^0 - E_{min})$ at which there is effectively no shift i.e. $\gamma_f(E_c^0 - E_{min}) = 0$. Then above equation reduces to

$$\sigma_{00} = \sigma_M \exp[(\gamma_c/k - GE_{min})] \quad (9)$$

If the shift in band edges $\gamma_c$ is known, then for such a value of $\sigma_{00}$ where $G=0$ (derived by extrapolation), one can obtain the value of $\sigma_M$. This information can further provide those values of $\sigma_0$ (from Eq. 6), where $\gamma_f = 0$, and where $\gamma_c = \gamma_f$, both very important positions for providing simplified information about the nature of carrier transport in the material. The quantity $E_{min}$ is a measure for the position of the DOS minimum within the mobility gap. In Fig. 4, we have plotted $G$ with $\sigma_{00}$ derived from the data of the above groups [shown in Fig 3(a & b)] and other groups from literature [case #8 (Ref.55), case #9 (Ref. 49]. The solid line shows the fit using Eq. (9). From here we have calculated the value of $\sigma_M$ (where $\gamma_f = \gamma_c$) $\approx$ 100 $(\Omega cm)^{-1}$, and found the minimum value of $E_{min}$ $\approx$ 0.61 eV which is a measure of the position of the DOS minimum within the gap. Using these values we find $\sigma_0 \approx 1.2 \times 10^3$ $(\Omega cm)^{-1}$ when $\gamma_f = 0$. Consolidating all these data and comparing with the analysis of the data in Fig. 2 (Lines 1 & 2), we find that our data is corroborative with these values that have been derived for a large number and variety of $\mu$c-Si:H materials, representative of the generic $\mu$c-Si:H system.

It follows from this study that a shift in the Fermi level of $\mu$c-Si:H material induced by any means (doping or any change in microstructure and the consequent DOS features) can give rise to an appearance of MNR behavior in the dc conductivity. Thus, the application of statistical shift model to the experimentally observed MNR parameters leads us to the information about some fundamental carrier transport parameters of the $\mu$c-Si:H material and a knowledge about the physical basis of MNR behavior, all with considerable simplicity.

## V. CONCLUSION

In conclusion, both MNR and anti MNR can be seen in the dark conductivity behavior of highly crystalline single phase undoped $\mu$c-Si:H material, depending on the microstructure and the correlative DOS features. We have shown the applicability of percentage fraction of constituent crystalline grains in fully crystallized $\mu$c-Si:H material in predicting electrical transport behavior. The transport mechanism in $\mu$c-Si:H has largely been a moot point in literature. Our study strongly indicates the presence of a band tail transport in $\mu$c-Si:H. We have shown that the statistical shift model can successfully explain both the MNR and anti MNR behavior in our material. Our assertions are further validated by the analysis of experimental $\mu$c-Si:H transport data available in literature. We have derived well-substantiated and generalized values of $E_{min}$, $\sigma_M$, and values of $\sigma_0$ and $E_a$ where $\gamma_f = 0$ and $\gamma_f = \gamma_c$, which hold true for the $\mu$c-Si:H system as a whole, and can further add to our understanding of the electrical transport in this heterogeneous material.

---

[1] W. Meyer and H. Neldel, Z. Tech. Phys. **18**, 588 (1937).
[2] W.B. Jackson, Phys. Rev. **B 38**, 3595 (1988).
[3] A. Yelon, B. Movaghar, and H.M. Branz, Phys. Rev. **B 46**, 12244 (1992).
[4] H. Overhof and W. Beyer, Philos. Mag. **B 47**, 377 (1983).
[5] B.G. Yoon, C. Lee, and J. Jang, J. Appl. Phys. **60**, 673 (1986).
[6] T. Drusedau and R. Bindemann, Phys. Stat. Sol. (**b**) **136**, K61 (1986).
[7] T. Drusedau, D. Wegener, and R. Bindemann, Phys. Stat. Sol. (**b**) **140**, K27 (1987).
[8] J. Stuke, J. Non-Cryst. Solids, **97-98**, 1 (1987).
[9] Minoru Kikuchi, J. Appl. Phys. **64**, 4997 (1988).
[10] S.R. Elliott, Physics of Amorphous Materials, 2nd ed., Longman Group UK Limited, England, (1990).
[11] E. Vallat-Sauvain, U. Kroll, J. Meier, N. Wyrsch, and A. Shah, J. Non-Cryst. Solids **266-269**, 125 (2000).
[12] M. Luysberg, P. Hapke, R. Carius, and F. Finger, Philos. Mag. **A 75**, 31 (1997).
[13] L. Houben, M. Luysberg, P. Hapke, R. Carius, F. Finger, and H. Wagner, Philos. Mag. **A 77**, 1447 (1998).
[14] G. Lucovsky, C. Wang, M.J. Williams, Y.L. Chen, and D.M. Maher, Mater. Res. Soc. Symp. Proc. **283**, 443 (1993).
[15] R. Vanderhaghen, S. Kasouit, J. D-Lacoste, F. Linu, and P. Roca i Cabarrocas, J. Non-Cryst. Solids **338-340**, 336 (2004).
[16] S. Kumar, R. Brenot, B. Kalache, V. Tripathi, R. Vanderhaghen, B. Drevillon, and P. Roca i Cabarrocas, Solid State Phenomena **80-81**, 237 (2001).
[17] P. Roca i Cabarrocas, S. Kasouit, B. Kalache, R. Vanderhaghen, Y. Bonnassieux, M. Elyaakoubi, and I. French, Journal of the SID **12**, 1 (2004), and references therein.
[18] S.K. Ram, D. Deva, P. Roca i Cabarrocas and S. Kumar, Thin Solid Films (2007) (in print).
[19] D. Azulay, I. Balberg, V. Chu, J.P. Conde, and O. Millo, Phys. Rev. **B 71**, 113304 (2005), and references therein.
[20] G. Lucovsky and H. Overhof, J. Non-Cryst. Solids **164-166**, 973 (1993).
[21] R. Fluckiger, J. Meier, M. Goetz, and A. Shah, J. Appl. Phys. **77**, 712 (1995).
[22] R. Brüggemann, M. Rojahn and M. Rösch, Phys. Stat. Sol. **166**, R11 (1998).
[23] H. Meiling and R.E.I. Schropp, Appl. Phys. Lett. **74**, 1012 (1999).
[24] M. Goerlitzer, N. Beck, P. Torres, U. Kroll, H. Keppner, J. Meier, J. Koehler, N. Wyrsch, and A. Shah, Mat. Res. Soc. Symp. Proc. **467**, 301 (1997).
[25] B. Yan, G. Yue, J.M. Owens, J. Yang, S. Guha, Appl. Phys. Lett. **85**, 1925 (2004).
[26] C.H. Lee, A. Sazonov, A. Nathan, and J. Robertson, Appl. Phys. Lett. **89**, 252101 (2006).
[27] S.K. Ram, Ph.D. thesis, I.I.T. Kanpur, India, (2006).
[28] Md. N. Islam and S. Kumar, Appl. Phys. Lett. **78**, 715 (2001).
[29] Md. N. Islam, A. Pradhan and S. Kumar, J. Appl. Phys. **98,** 024309 (2005).
[30] S.K. Ram, P. Roca i Cabarrocas and S. Kumar, Thin Solid Films (2007) (in print).
[31] N.F. Mott and E.A. Davis, Electronic Processes in Non Crystalline Materials, 2nd ed., Oxford University Press, (1979).
[32] M. Cuniot and Y. Marfaing, Philos. Mag. **B 5**, 291 (1988).
[33] H. Mimura and Y. Hatanaka, Appl. Phys. Lett. **50**, 326 (1987).
[34] X. Xu, J. Yang, A. Banerjee, S. Guha, K. Vasanth, and S. Wagner, Appl. Phys. Lett. **67**, 2323 (1995).
[35] S. Hamma and P. Roca i Cabarrocas, Appl. Phys. Lett. **74**, 3218 (1999).
[36] P. Irsigler, D. Wagner and D.J. Dunstan, J. Phys. **C 16**, 6605 (1983).
[37] M.J. Powell and S.C. Deane, Phys. Rev. **B 48**, 10815 (1993).
[38] G. Schumm, Phys. Rev. **B 49**, 2427 (1994).
[39] A. Merazga, H. Belgacem, C. Main, and S. Reynolds, Solid State Commun. **112**, 535 (1999).
[40] P.G. Lecomber, G. Willeke and W.E. Spear, J. Non-Cryst. Solids **59-60**, 795 (1983).
[41] K. Shimakawa, J. Non-Cryst. Solids **266-269**, 223 (2000).
[42] J. Kocka, A. Fejfar, P. Fojtik, K. Luterova, I. Pelant, B. Rezek, H. Stuchlikova, J. Stuchlik, and V. Svrcek, Sol. Energy Mat.& Sol. Cells **66**, 61 (2001), and references therein.
[43] J. Kocka, H. Stuchlikova, J. Stuchlik, B. Rezek, T. Mates, V. Svrcek, P. Fojtik, I. Pelant, and A. Fejfar, J. Non-Cryst. Solids **299-302**, 355 (2002), and references therein.
[44] J.Y.W. Seto, J. Appl. Phys. **46**, 5247 (1975).
[45] M.V. Garcia-Cuenca, J.L. Morenza and J. Esteve, J. Appl. Phys. **56**, 1738 (1984).
[46] S.K. Ram, S. Kumar, R. Vanderhaghen, B. Drevillon, and P. Roca i Cabarrocas, Thin Solid Films **511-512**, 556

(2006).
[47] S.K. Ram, P. Roca i Cabarrocas and S. Kumar, J. Non-Cryst. Solids **352**, 1172 (2006).
[48] S.K. Ram, P. Roca i Cabarrocas and S. Kumar, Thin Solid Films (2007) (in print).
[49] Y. Lubianiker and I. Balberg, Phys. Rev. Lett. **78**, 2433 (1997).
[50] R.W. Collins, A.S. Ferlauto, G.M. Ferreira, C. Chen, J. Koh, R.J. Koval, Y. Lee, J.M. Pearce, and C.R. Wronski, Sol. Energy Mat.& Sol. Cells **78**, 143 (2003).
[51] D. He, N. Okada, C.M. Fortmann, and I. Shimizu, J. Appl. Phys. **76**, 4728 (1994).
[52] X. Wang, Y. Bar-Yam, D. Adler, and J.D. Joannopoulos, Phys. Rev. **B 38**, 1601 (1988).
[53] S.Y. Myong, O. Shevaleevskiy, K.S. Lim, S. Miyajima, and M. Konagai, J. Appl. Phys. **98**, 054311 (2005).
[54] S. Y. Myong, K. S. Lim, M. Konagai, Appl. Phys. Lett. **88**, 103120 (2006).
[55] D. Han, G. Yue, J. D. Lorentzen, J. Lin, H. Habuchi, and Q. Wang, J. Appl. Phys. **87**, 1882 (2000).